\documentclass[fleqn,usenatbib]{mnras}

\usepackage[T1]{fontenc}
\usepackage{ae,aecompl}

\usepackage{graphicx}	
\usepackage{amsmath}	
\usepackage{amssymb}	

\title[gas accretion onto galaxies]{Gas accretion onto galaxies and Kelvin-Helmholtz turbulence} 
\author[Goldman and Fleck]{ Itzhak Goldman$^{1, 2}$ \thanks{E-mail:goldman@afeka.ac.il} and Robert Fleck$^3$   \\
$^1$ Department of Physics, Afeka College, Tel Aviv, Israel \\
$^2$ Department of Astrophysics, Tel Aviv University, Tel Aviv, Israel \\ 
$^3$ Department of Physical Sciences, Embry-Riddle Aeronautical University, Daytona Beach, Florida, USA} 
 \pubyear{2022}

\begin{document}
\label{firstpage}
\pagerange{\pageref{firstpage}--\pageref{lastpage}}
\maketitle 
\begin{abstract}
 
Continued star formation over the lifetime of  a galaxy suggests that low metalicity gas is steadily flowing in from the circumgalactic medium.   Also,  cosmological   simulations of  large-scale structure formation   imply that   gas is accreted  onto galaxies from the halo inside which they formed. Direct observations are difficult, but in   recent years   observational indications  of gas inflows from a circumgalactic medium were obtained.  
Here  we suggest an   {\it indirect} observational probe: looking for   large-scale (exceeding few kpc) turbulence caused by the accretion. As a specific example    we consider an accretion flow coplanar with the galaxy disk, and argue that  Kelvin-Helmholtz turbulence will be generated. We employ a semi-analytic model of turbulence and derive the expected turbulence power spectrum. The latter turns out to be of a  distinctive shape that can be compared with observational power spectra.  As an illustrative example we use parameters of the Milky Way galaxy.  
\end{abstract}


\section{Introduction}  

The observational evidence that star formation goes on continuously in galaxies suggests that  there is 
 fresh gas supply from   the circumgalactic medium (CGM); (see e.g. \cite{ Elmegreen16, B-H+17}. 
Simulations of  cosmological large-scale structure  formation  support this picture;  e.g. \citep{Keres+2005}. For a recent review on the physics, astrophysics and observational status of the CGM  see \cite{Tumlinson+17}.
  
  Direct observations are difficult; nevertheless, in   recent years observation were made   suggesting the existence of such an accretion  \citep{Elmegreen+2014, Elmegreen+2016,  Kacprzak2017, Vulcani+2018, Martin+2019, Zabl+2019, Das+2020, Luo+2021, Ianjamasimanana+2022}. 
 
Recently, observations carried out for disk galaxies at  a redshift $z\sim 0.2$ \citep{Ho+2017,Ho+2019,Martin+2019,Ho+Martin2020}  and  at redshift $z\sim 1$ \citep{Zabl+2019} 
 indicate the existence of cold gas inflowing coplanar with  the disk  in the near CGM (30--80 kpc) corotating 
with the  disk. The estimated inflow radial velocities are in the range of (20--60)~ km/s.  
   A very recent simulation by \cite{Trapp+2022} conducted for  a MIlky Way (MW)-type  galaxy obtained results consistent with
these   observations, finding that the deviation of accreting gas from the circular velocity of the  galaxy disk gas is quite small and that there is a radial inflow velocity  $\sim 40\  $~ km/s.

In the present  paper we examine a possible  observational signature of  
   disk-plane  accretion of   CGM  gas.
  We find that such an accretion could generate    Kelvin-Helmholtz (KH) 
 turbulence at the disk outskirts. 
  Specifically, we address the case in which the inflowing gas is cool and rotates in the same direction as that of the galactic disk. 
 as supported by the observations and simulations; the parameters of the  MW  galaxy are used as an illustrative  example.  
  
  The resulting turbulence is large scale ($\sim 10 \ $~kpc)  with a distinctive  velocity power spectrum. There is a range of wavenumbers in which the power spectrum has a logarithmic slope of $\sim -1.1$ which is quiet distinct from the Kolmogorov slope of $-5/3$ or $-2$ corresponding to compressible turbulence. In order to test the sensitivity of the results to  values of the adopted  parameters, we repeat the computation with others  that are consistent with the observations.  
While the   largest scale of the turbulence  and the  turbulent velocity do depend on the adopted values of the  parameters used, the power spectrum retains its particular shape.

 \section{Kelvin-Helmholtz Instability}
 
 The  KH  instability (KHI) arises when two fluids with different velocities and (usually) densities
 form an interface perpendicular to a gravitational field   \citep{chandra61}. It is   manifested in the case of  wind flowing over a body of water, and also in various astrophysical settings   \citep{Fleck83, Fleck84, Fleck89, Gomez+Ostriker2005, Mandelker+2016,Fleck2020}.   In this paper we apply  the linear growth rate of the instability in a semi-analytic model of turbulence \citep{CGM96}. The latter  re-normalizes the  growth rate self-consistently taking into account non-linear turbulence. The turbulence model provides the spatial power spectrum of the turbulent velocity field.

 The linear growth rate of the KHI   is  given  by \cite{chandra61}
 \begin{eqnarray}
 n_s = 
 \sqrt {\alpha_1\alpha_2  k_{||}^2 V_{rel}^2  - \frac{B(k)^2 }{2\pi (\rho_1 + \rho_2)}  k_{||}^2    - g k (\alpha_1 -\alpha_2)},
  \end{eqnarray}
   where  $\alpha_1= \frac{\rho_1}{\rho_1 + \rho_2}$ and $\alpha_2= \frac{\rho_2}{\rho_1 + \rho_2}$  
 with $\rho_1$ and $\rho_2$ the mass density of the disk and the accreted gas, $g$ the vertical gravitational acceleration, $V_{rel}$ is the     relative velocity between the accreted gas and the galactic disk gas, $B(k)$  the absolute value of the magnetic field at a spatial scale $2\pi/k$, $k$  the absolute value of the wavenumber, and $k_{||}$   the component of the wavenumber  parallel to the relative velocity.

 The condition for instability in the presence of rotation was obtained by \cite{Huppert68}: 
  \begin{equation}
  k_{||}^2V_{rel}^2\geq 4 \Omega^2, 
    \end{equation}
     where $\Omega$ is the   angular velocity.
   As we shall see, this condition is satisfied with a wide margin, and therefore the growth rate given by equation (1) is appropriate here.
   
     \section{The astrophysical setting considered} 
 We consider the case of  a disk galaxy similar to the MW. We adopt  parameters consistent with  the observations of
 \cite{Ho+2017}, \cite{Ho+2019}, \cite{Martin+2019}, \cite{Ho+Martin2020}, \cite{Zabl+2019}, and the simulations of \cite{Trapp+2022}.
  We  consider a cold CGM gas  that at a galactic radius $R=40$~ kpc has approximately the same rotational velocity as that of the   disk. We thus adopt a relative velocity which is radial (inward); $V_{rel} = 40$~ km/ s. At the galaxy outskirt we take
$\rho_1= 1.67 \times 10^{-25}$~g  cm$^{-3}$ corresponding to a number density of $0.1$~ cm$^{-3}$. The inflowing gas density at the above radius was taken to be $\rho_2= 0.055\rho_1= 9.2 \times 10^{-27}$ ~g  cm$^{-3}$. This yields $\alpha_1=0.95, \ \alpha_2= 0.05$.

These parameters imply a mass accretion rate for an angular extension of the inflowing gas, $ 0 < \beta< 2\pi$, and a scale height $H=330$~pc

\begin{equation}
\dot M=  \beta 2 H R \rho_2 V_{rel} = 0.13  \beta \ {\rm M_{\odot}\  yr^{-1}}
\end{equation}
 which is quite reasonable.

\subsection{The vertical acceleration g}
 
The vertical acceleration is the   sum of the vertical components of the galactic acceleration and the self gravity of the disk:

\begin{eqnarray}
\label{g}
g= g_{gal}+  g_{self},  \\
 \nonumber
{\rm where}\ g_{gal}=\frac{ v_{rotation}^2 }{R}\frac{H}{R} \ \ , \ {\rm and} \  g_{self}= 2\pi G \Sigma
  \end{eqnarray}
 where $\Sigma$ is the surface mass density of the disk.
 At $R=40$~kpc we adopt the parameters from  \cite{Sofue2013}: $v_{rotation}= 158\ {\rm km /s, \ \ {\rm and }\
  \Sigma= 1 \  M_{\odot}  \ pc^{-2}}$.
 These imply $g_{gal}= 1.8\times 10^{-11}{\rm  cm\ s^{-2}}$ and $ g_{self}=  8.8\times 10^{-11}{\rm cm\ s^{-2}}$, yielding $g = 1.06\times 10^{-10} {\rm cm\ s^{-2}}$.

  \subsection{Random galactic magnetic field}
  
Many observations were made with  the aim of revealing the nature of the  galactic random magnetic field.   \cite{Rand+Kulkarni89}, \cite{Ohno+Shibata93}, \cite{Han+2004}, and \cite{Han2017} concluded that measurements of the rotation  and dispersion measures of pulsars can be well represented by a random field with coherence length ("cell")  
  $L_B$ and field strength   $ B_0$. As a result, the average field along a given line of sight of length $L=2\pi/k$  due to the  randomly oriented in-cell fields is

 \begin{equation}
 B(k)= B_0\sqrt{\frac{k}{k_B}},\ \  k\leq k_B, \   \ {\rm where}\ k_B= \frac{2 \pi}{L_B   }. 
    \end{equation}
  Substituting this to equation (1) yields
 
\begin{eqnarray}
  n_s = 
  \sqrt {\alpha_1\alpha_2  k_{||}^2 V_{rel}^2   - \frac{B_0^2}{2 \pi (\rho_1 + \rho_2)}\frac{  k_{||}^3 }{k_B}- g k (\alpha_1 -\alpha_2)}.
 \end{eqnarray}
We adopt $B_0= 1\  \mu$~G and $k_B=  2\pi /(200\ {\rm pc})$.  
Substitution of  the numerical values shows that $n_s(k)$ is real and positive for wavenumbers  
 $ 1.31  \times 10^{-22}{\rm cm^{-1}} < k  <  8.51 \times 10^{-21}{\rm cm^{-1}}$ corresponding to spatial scales    
 $ \   0.24\ {\rm kpc}        < 2\pi/k< 15.5\ {\rm kpc}$.

\section{turbulence model}
 Extending  the earlier  turbulence models of    \cite{CG85} and \cite{CGC87}, we employ the model of \cite{CGM96}  
      to derive  the expected power spectrum of the turbulence.  
The model is formulated as an integral equation which  represents a balance   between the net rate of energy input to  the turbulence in the wavenumber range $(k_0-k)$ and the   rate of the energy cascaded to all scales smaller than $k$,
  
 \begin{eqnarray}
    \int_{k_0}^k  n_s(k')dk' = y(k) \nu_t(k), \\ 
{\rm where}  \ \ \ \ y(k)= \int_{k_0}^k F(k') k'^2 dk'. 
    \end{eqnarray}
      $F(k)$  is the power spectrum of the turbulent velocity, $y(k)$ is the  $k$-space mean square  vorticity at wavenumber $k$, $n_s(k)$ is the net rate controlling the energy input from the source at $k$ incorporating the rate of energy dissipation by molecular viscosity, and    
      $\nu_t(k)$ is the turbulent kinematic viscosity at wavenumber $k$ exerted by 
 all the eddies with   wave number larger than $k$. Here, $k_0$ is the wavenumber corresponding to the largest scale of the turbulence. 
  The turbulent viscosity at wavenumber $k$ is defined by
\begin{equation}
\nu_t(k)= \int_k^{\infty}\frac{F(k') }{n_c^*(k')} dk'.
     \end{equation}
    Here, $n_c^*(k)$ is the rate controlling the eddy nonlinear correlation (heuristically the rate of the eddy breakup at $k$).
 
By differentiating equation (7) it is possible to obtain a   rate  equation:
 
 \begin{equation}
  n_s(k) + \frac{y(k)}{n_c^*(k)}= \nu_t (k) k^2  =\gamma n_c(k).
 \end{equation}
  The second term on the left hand side is the rate controlling the energy cascaded from  all spatial scales  
larger   $2\pi/k$. The right hand side is the rate at which energy is transferred to the smaller scales, with 
$\gamma$  a positive dimensionless constant  which was fitted by \cite{CGM96}   to produce a Kolmogorov inertial power spectrum, yielding $\gamma= 0.088\left(K_c/1.5\right)^{-3}$. Here $K_c$ denotes the so-called Kolmogorov constant.

     The eddy correlation rate is modeled as
     \begin{equation}
      \gamma n_c^*(k)= \left([\gamma n_c(k)]^{1/2}+[n_s(k)]^{1/2}\right)^2,
     \end{equation}
  and is dependent on both   the turbulent viscosity and on   $n_s(k)$.
For a given $n_s(k)$, the solution of the coupled equations yields the power spectrum $F(k)$.
The largest spatial scale corresponding to the smallest wavenumber $k_0$  is obtained from   
\begin{equation}
\frac{d}{dk}\left(\frac{n_s(k)}{ k^2}\right)_{k_0} = 0,
\end{equation}   
 and at this scale $F(k_0)=0$.

By combining equation~(10) and ~(11), one obtains an algebraic expression of $y(k)$   in term of $n_s(k)$ and  $n_c(k)$:
\begin{equation}
y(k) = \gamma^{-1}\left(   \gamma n_c(k) - n_s(k)\right)\left([\gamma n_c(k)]^{1/2}+[n_s(k)]^{1/2}\right)^2.
\end{equation}
Differentiating equation (9)  gives
\begin{eqnarray}
F(k)= -\left([\gamma n_c(k)]^{1/2}+[n_s(k)]^{1/2}\right)^2 \frac{d }{dk}\left(\frac{n_c(k)}{k^2}\right) 
 \end{eqnarray}
 which upon using the relation $ dy(k) /dk = F(k) k^2$ leads to
 \begin{eqnarray}
\frac{dy(k)}{dk} = - k^2  \left([\gamma n_c(k)]^{1/2}+[n_s(k)]^{1/2}\right)^2 \frac{d }{dk}\left(\frac{n_c(k)}{k^2}\right),
\end{eqnarray}
From equations ( 13) and (15) one obtains a first-order differential equation for $n_c(k)$. The initial condition is 
$ \gamma n_c(k_0)=n_s(k_0)$.  Once $n_c(k)$ has been solved, equation (14) determines $F(k)$.

\section{Computed power spectrum and the  turbulent velocity}   
    
 We take $n_s(k)$ to be the linear growth rate given by equation (6). As has been argued by \cite{CGM96}, the self-consistent formulation of the eddy correlation timescale effectively modifies the growth rate and makes it dependent also on the turbulence.
We obtain the velocity power spectrum and the  value of the turbulent velocity for two sets of parameters  that are within the range suggested by the observations and simulations. It is of interest to identify
which   features  are independent of the precise  values of the parameters.

\subsection{Parameters set 1}
For the parameters listed in the previous sections,   equation (12) yields
\begin{eqnarray}
k_0=  1.96\times 10^{-22}\ {\rm cm^{-1}   \ and }\\
\nonumber
 L_0= 2  \pi/k_0 = 10.4 \ {\rm \ kpc}. 
\end{eqnarray}
The smallest scale $2\pi /k_f$  for which $n_s(k_f)=0$ is  $240$~pc. Note that
   $L_0$, the largest scale of the turbulence, is about $2/3$ of the largest scale at which $n_s>0$. 
The formal solution of the power spectrum for $ k<k_0$ is negative and thus no turbulence exists in this range even though 
  $n_s$ is positive there. The growth rate at $k=k_0$ is
 \begin{equation}
   n_s( k_0)= 9.9\times 10^{-17}s^{-1} = \frac{1}{3.3 \times 10^8\ yr}.
  \end{equation}
 The growth rate $n_s(k)$ and the power spectrum $F(k)$ are shown in Fig. 1 and Fig. 2, respectively. The power spectrum  rises from $F(k_0)=0$, reaches a maximum at $k \simeq 3 k_0 $ corresponding to a spatial scale of $\simeq 3.3$~kpc and then declines with a logarithmic slope that changes from -1 to -1.2 as $k$ increases.
 
 The   turbulent velocity is
  \begin{equation}
  v_{turb}= \sqrt{\int _{k_0}^{k_f} F(k) dk}= 32.6\gamma^{-1/2}\ {\rm km/s}.
  \end{equation}
  
 \subsection{Parameters set 2}

Here the computation is repeated for $V_{rel} = 80$~ km/s, which is at the high end of the observational and simulated  values.  In order not to obtain an unrealistically high mass accretion rate, a lower value of $\rho_2 $ is adopted: $\rho_2= 3.06\times 10^{-27}\ {\rm g\ cm^{-3}}$.
Thus, $\alpha_1= 0.982$ and $\alpha_2= 0.018$. The mass accretion rate for these parameters is 
  is 
\begin{equation}
\dot M=  \beta 2 H R \rho_2 V_{rel} = 0.1  \beta  M_{\odot}\ {\rm yr^{-1}},
\end{equation}
similar to that obtained   for the previous parameter  set. The largest turbulence scale is  larger (reflecting the larger value of $ V_{rel}$): 
\begin{eqnarray}
 k_{02}=  1.4\times 10^{-22}\ {\rm cm^{-1}} \\
\nonumber
 L_{02}= 2  \pi/k_0 = 14.5 \ {\rm kpc}. 
 \end{eqnarray}
 The smallest scale is $165$~pc. Since the coherence length of the magnetic field was taken as $200$~pc, the power spectrum smallest scale was taken as  $200$~pc corresponding to a relative wave number  $k/k_ 0= 72.5 $.
 The  turbulent velocity for these parameters is
 \begin{equation}
  v_{turb}= \sqrt{\int _{k_0}^{k_f} F(k) dk}= 46.5\gamma^{-1/2}\ {\rm km/s}.
  \end{equation} 
  The growth rate $n_s(k)$ and the power spectrum $F(k)$ are shown in Fig. 3 and Fig. 4, respectively. 
We note that the power spectrum   has the same  distinctive shape as in set 1 despite the different values of the spatial scales and of the  turbulent velocity.
 
\section{discussion and conclusions}
 The power spectrum of the turbulence (Fig. 2 and Fig. 4) has   a unique shape. It spans wavenumbers corresponding to spatial scales in the range ($  (10.4 -0.22)\ {\rm  kpc}$ in case 1. and  $  (14.5 -  0.20) \ {\rm  kpc}$ for the parameters of case 2. It is zero for the largest scale, rises and then declines, almost with a constant logarithmic slope.   For intermediate   spatial   scales the power spectrum has a logarithmic slope of $-1.1$, which turns gradually to a value of -2 for smaller scales.
 
 While the  regions where the power spectrum has a power-law dependence on the wavenumber resembles 
  an inertial range, this not the case,   since  $n_s(k)$  is not zero in this range. Actually, the  latter fact is the reason for the shallow decline of the power spectrum (compared to the Kolmogorov spectral logarithmic slope of $-5/3$).
   For $k$-values larger than the maximal $k$ (where the energy input vanishes) there may exist an inertial range proportional to $k^{-5/3}$ provided  that the microscopic viscosity can be neglected.  

The turbulent  velocity is $ 32.6\ \gamma^{-1/2}  {\rm km/s}  $ for case 1.  and  $ 46.5\ \gamma^{-1/2} {\rm km/s} $ for case 2. For the value of $\gamma\simeq 0.9 $  obtained by \cite{CGM96} the turbulent velocities are quiet large. This is due to the 
wide range and slow decline of the power spectrum.

We note that the time scale $n_s(k)^{-1}$ is shorter than the rotation period at $R=40\ $~kpc. In addition, the timescales that
characterize the turbulence: $n_c(k) ^{-1}$ and    $n_c^*(k) ^{-1}$ are much shorter.

Obviously, the  numerical values depend on the assumed radial velocity and on the value of
$ \rho_2 $.  The range of the spatial scales of the turbulence  changes as well as the normalization of the power spectrum and the turbulent velocity. However, the unique shape of the power spectrum is  unchanged, a feature that identifies the   turbulence as KH.  
 
  Observation  of a  fluctuating velocity field at the outskirts of  a galaxy having  a large-scale power spectrum with 
 the shape obtained here could serve as   indirect observational evidence for gas accretion from the CGM. 
 
\begin{figure} 
 \centerline{\includegraphics[scale=0.5]{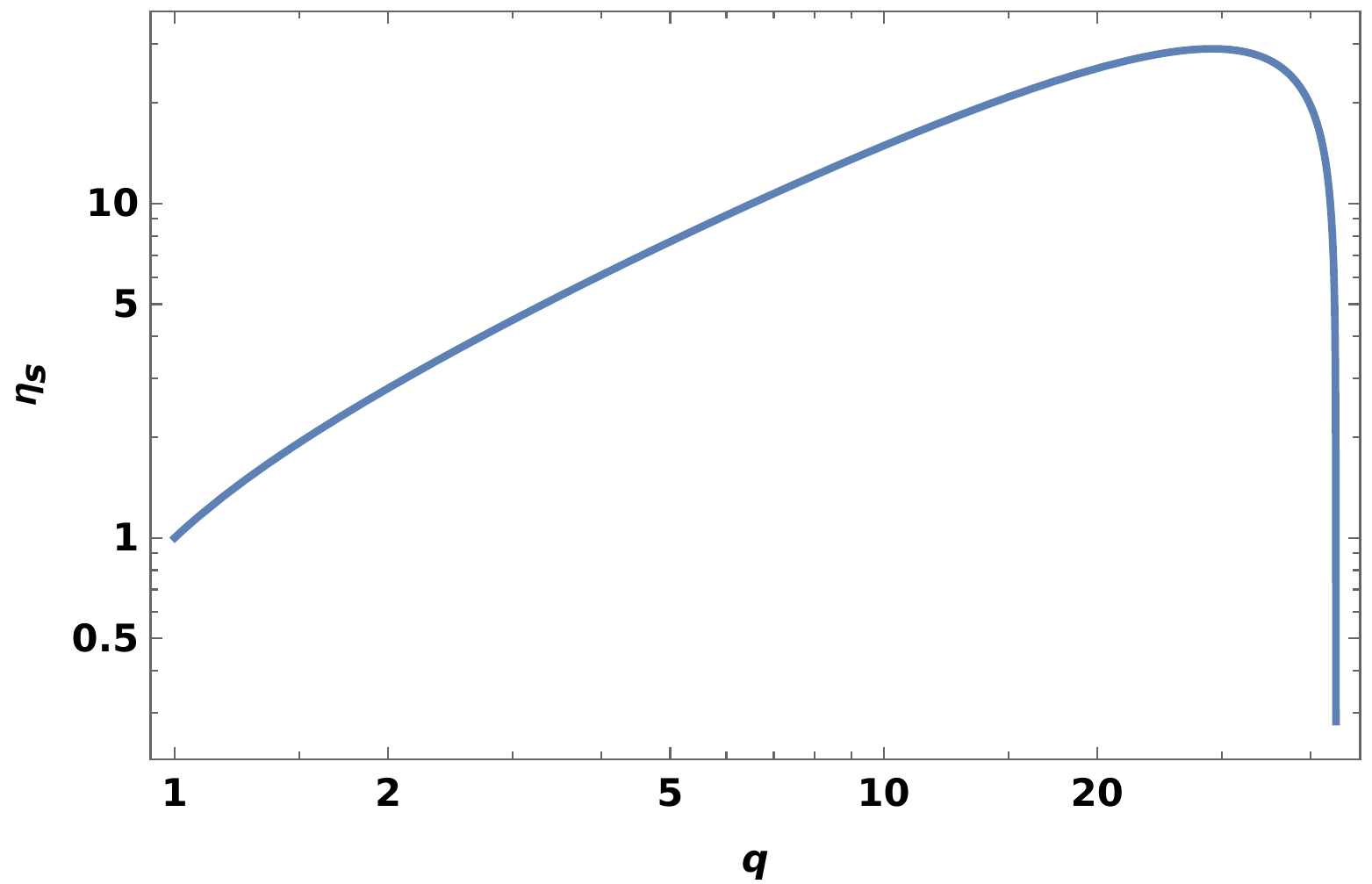}}
    \caption{Set 1:The dimensionless growth rate $\eta_s  \equiv n_s(k)/n_s(k_0)$ as a function of the dimensionless wavenumber $q\equiv k/k_0$.} 
  \end{figure}
  
 \begin{figure} 
 \centerline{\includegraphics[scale=0.5 ]{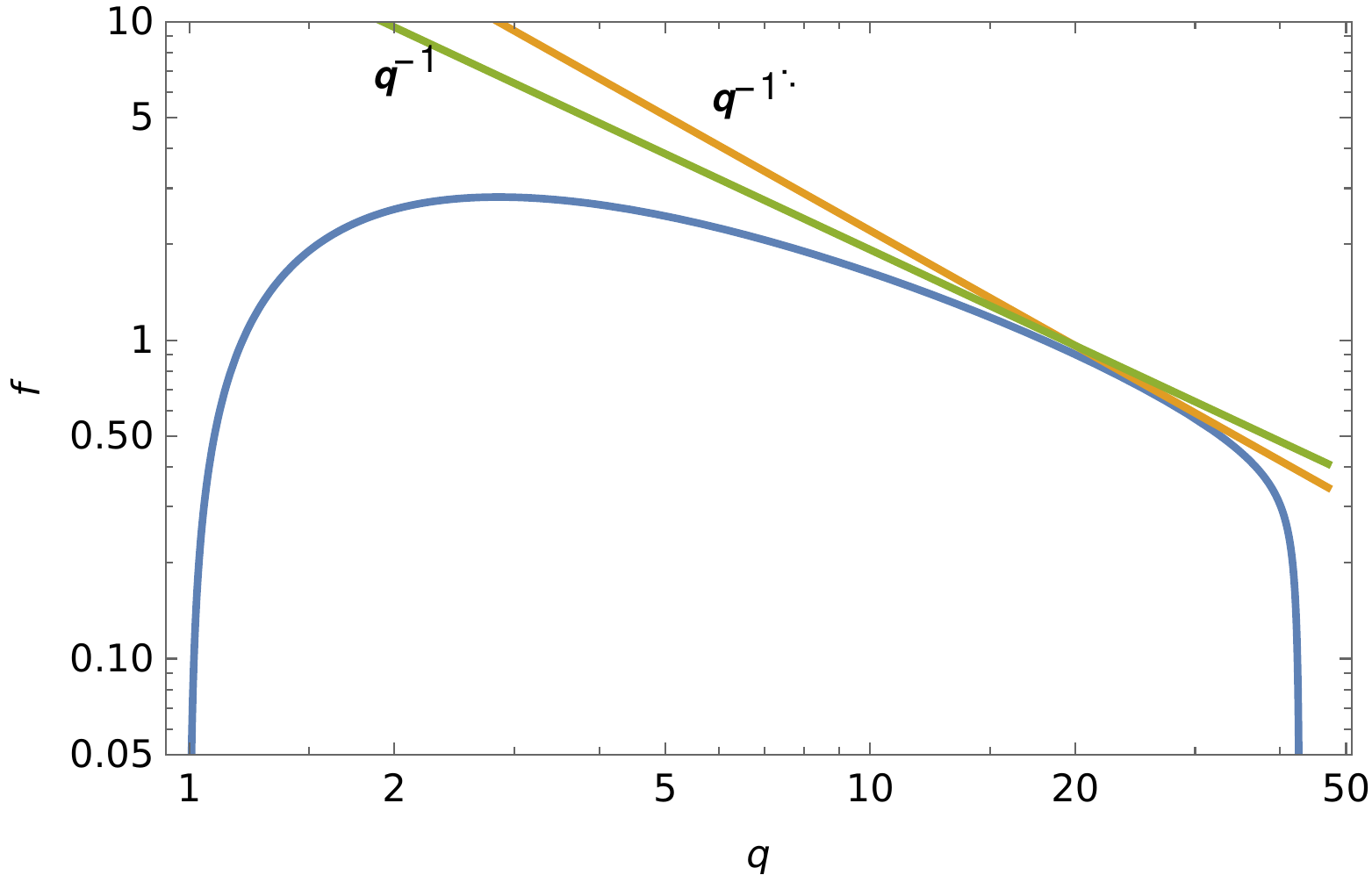}}
  \caption{Set 1: The dimensionless power spectrum \\$f \equiv F(k)\left(n_s(k_0)/k_0\right)^{-2}\gamma$   as a function of  dimensionless wavenumber $q\equiv k/k_0 $. The Orange line has a logarithmic slope of $-1.2$; the Green line has a logarithmic slope of $-1$. } 
 \end{figure}
 
\begin{figure} 
 \centerline{\includegraphics[scale=0.5]{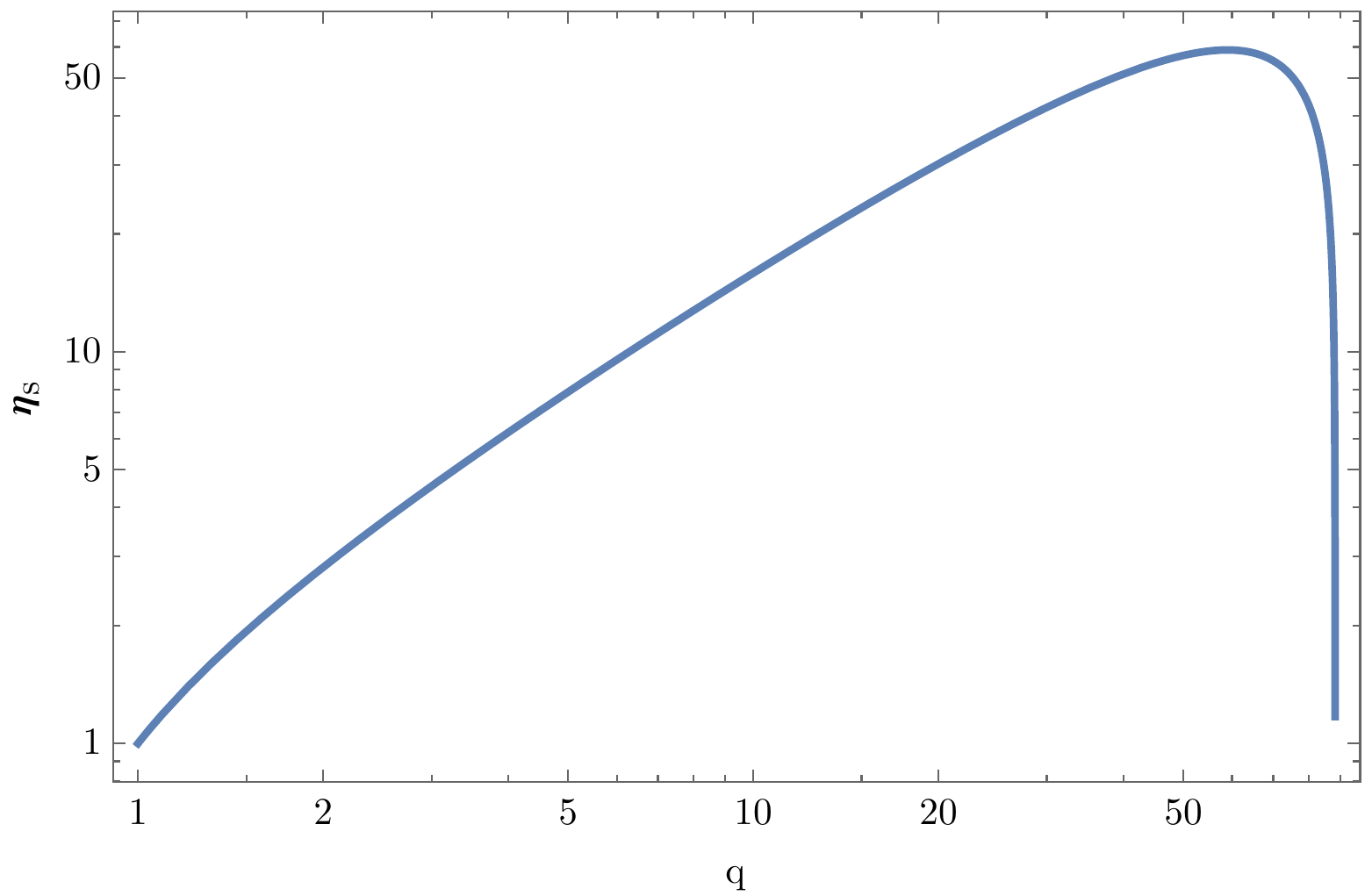}}
    \caption{Set 2: the dimensionless growth rate $\eta_s  \equiv n_s(k)/n_s(k_0)$ as a function of the dimensionless wavenumber $q\equiv k/k_0$.}
  \end{figure}
 \begin{figure} 
 \centerline{\includegraphics[scale=0.5 ]{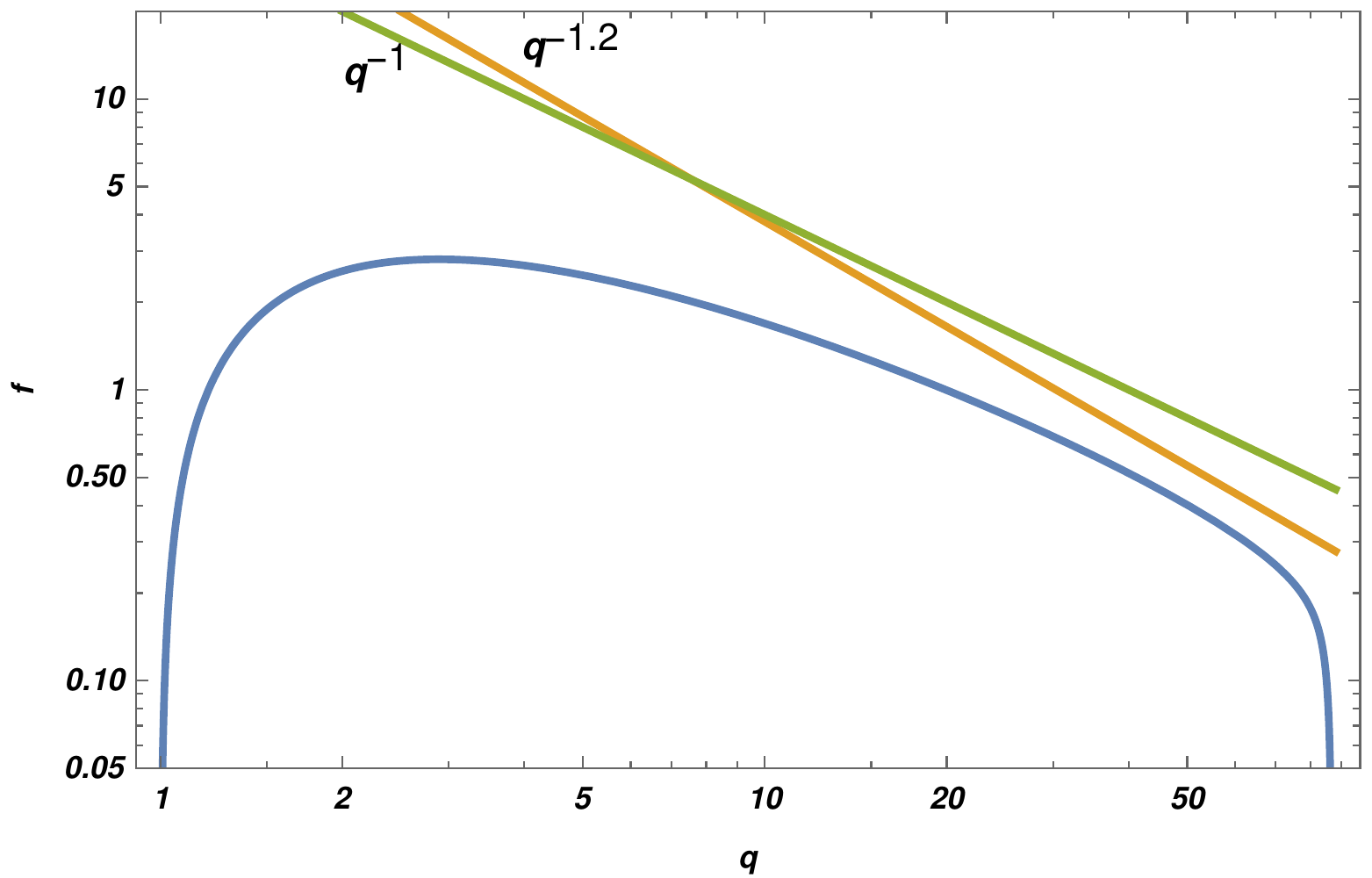}}
  \caption{Set 2:  The dimensionless power spectrum \\ $f\equiv F(k)~\left(n_s(k_0)/k_0\right)^{-2}\gamma$   as a function of  dimensionless wavenumber $q\equiv k/k_0 $. The Orange line has a logarithmic slope of $-1.2$; the Green line has a logarithmic slope of $-1$.  } 
 \end{figure} 
 \section*{Acknowledgements}
Itzhak Goldman thanks Afeka College,  as well as the  Astrophysics department  of Tel Aviv University.

\section{data availability}
This is  a theoretical paper. No specific data is analyzed.

\label{lastpage}

\end{document}